
\input harvmac
\baselineskip 14pt
\input epsf
\magnification\magstep1
\parskip 6pt
\newdimen\itemindent \itemindent=32pt
\def\textindent#1{\parindent=\itemindent\let\par=\resetpar%
\indent\llap{#1\enspace}\ignorespaces}

\let\oldpar=\par
\def\resetpar{\oldpar\parindent=20pt\let\par=\oldpar}

\font\ninerm=cmr9 \font\ninesy=cmsy9
\font\eightrm=cmr8 \font\sixrm=cmr6
\font\eighti=cmmi8 \font\sixi=cmmi6
\font\eightsy=cmsy8 \font\sixsy=cmsy6
\font\eightbf=cmbx8 \font\sixbf=cmbx6
\font\eightit=cmti8
\def\eightpoint{\def\rm{\fam0\eightrm}
  \textfont0=\eightrm \scriptfont0=\sixrm \scriptscriptfont0=\fiverm
  \textfont1=\eighti  \scriptfont1=\sixi  \scriptscriptfont1=\fivei
  \textfont2=\eightsy \scriptfont2=\sixsy \scriptscriptfont2=\fivesy
  \textfont3=\tenex   \scriptfont3=\tenex \scriptscriptfont3=\tenex
  \textfont\itfam=\eightit  \def\it{\fam\itfam\eightit}%
  \textfont\bffam=\eightbf  \scriptfont\bffam=\sixbf
  \scriptscriptfont\bffam=\fivebf  \def\bf{\fam\bffam\eightbf}%
  \normalbaselineskip=9pt
  \setbox\strutbox=\hbox{\vrule height7pt depth2pt width0pt}%
  \let\big=\eightbig  \normalbaselines\rm}
\catcode`@=11 %
\def\eightbig#1{{\hbox{$\textfont0=\ninerm\textfont2=\ninesy
  \left#1\vbox to6.5pt{}\right.\n@@space$}}}
\def\vfootnote#1{\insert\footins\bgroup\eightpoint
  \interlinepenalty=\interfootnotelinepenalty
  \splittopskip=\ht\strutbox %
  \splitmaxdepth=\dp\strutbox %
  \leftskip=0pt \rightskip=0pt \spaceskip=0pt \xspaceskip=0pt
  \textindent{#1}\footstrut\futurelet\next\fo@t}
\catcode`@=12 %
\def \b1{{\bf 1}}
\def \d{{\rm d}}
\def \de{\delta}
\def \si{\sigma}

\def \pr{\partial}
\def \d{{\rm d}}
\def \tr{{\rm tr }}
\def \bY{{\bar{Y}}}
\def \ta{{\tilde a}}
\def \tbe{{\tilde \beta}}
\def \hg{{\hat g}}
\def \hy{{\hat y}}
\def \hY{{\hat Y}}
\def \hbY{{\hat \bY}}

\def \rO{{\rm O}}

\def \half{{\textstyle {1 \over 2}}}

\def \quar{{\textstyle {1 \over 4}}}

\def \ts{ \textstyle}

\def \ep{\epsilon}

\font \bigbf=cmbx10 scaled \magstep1
\font \bigit=cmti10 scaled \magstep1

\lref\Shif{M. Shifman, Non-Perturbative Dynamics in Supersymmetric Gauge
Theories, lectures at the 1996 Trieste Summer School, hep-th/9704114\semi
M.E. Peskin, Proceedings of the 1996 Boulder Theoretical Advanced Study 
Institute, SLAC-PUB-7393, hep-th/9702094.}
\lref\Cardy{J.L. Cardy, {Phys. Lett.} {B215} (1988) 749.}
\lref\cone{H. Osborn, {Phys. Lett.} {B214} (1988) 555.}
\lref\JO{H. Osborn, Phys. Lett. B222 (1989) 97\semi
I. Jack and H. Osborn, {Nucl. Phys.} {B343} (1990) 647.}
\lref\JJ{I. Jack, D.R.T. Jones and C.G. North, Nucl. Phys. B473 (1996) 308,
hep-ph/9603386.}
\lref\JJN{I. Jack, D.R.T. Jones and C.G. North, Nucl. Phys. B486 (1997) 479,
{hep-ph/9609325}.}
\lref\Weyl{H. Osborn, Nucl. Phys. B363 (1991) 486.}
\lref\Cap{A. Cappelli, D. Friedan and J.I. Latorre, {Nucl. Phys.}
{B352} (1991), 616.}
\lref\hughtwo{J. Erdmenger and H. Osborn, Nucl. Phys. {B483} (1997)
431; hep-th/9605009.}
\lref\hughone{H. Osborn and A. Petkou,
    Ann. Phys. {231} (1994) 311, hep-th/9307010.}
\lref\Bast{F. Bastianelli, Phys. Lett. B369 (1996) 249, hep-th/9511065.}
\lref\Gris{
D. Anselmi, D.Z. Freedman, M.T. Grisaru and A.A. Johansen, Phys. Lett. B394
(1997) 329.}
\lref\Anone{D. Anselmi, D.Z. Freedman, M.T. Grisaru and A.A. Johansen, 
preprint, hep-th/9708042.}
\lref\Antwo{D. Anselmi, J. Erlich,  D.Z. Freedman and A.A. Johansen, 
preprint, hep-th/9711035.}
\lref\NSVZ{V. Novikov, M.A. Shifman, A.I. Vainshtein and V. Zakharov,
Nucl. Phys. B229 (1983) 381\semi
M.A. Shifman and A.I. Vainshtein, Nucl. Phys. B277 (1986) 456\semi
V. Novikov, M.A. Shifman, A.I. Vainshtein and V. Zakharov, Phys. Lett. B166
(1986) 329\semi
M.A. Shifman, A.I. Vainshtein and V. Zakharov, Phys. Lett. B166 (1986) 334.}
\lref\LO{J.I Latorre and H. Osborn, Nucl. Phys. B511 (1998) 737, 
hep-th/9703196.}
\lref\Hath{S.J. Hathrell, {Ann. Phys.} {139} (1982) 136, {142} (1982) 34\semi
M.D. Freeman, {Ann. Phys.} {153} (1984) 339.}
\lref\grad{ D.J. Wallace and R.K.P. Zia, {Phys. Lett.} {48A} (1974) 327; 
{Ann. Phys.} {92} (1975) 142\semi
B.P. Dolan, Mod. Phys. Lett. 8 (1993) 3103.}
\lref\Kogan{I.I. Kogan, M. Shifman and A. Vainshtein, Phys. Rev. D53 (1996)
4526.}
\lref\Jack{I. Jack, Nucl. Phys. B253 (1985) 323.} 
\lref\Sei{N. Seiberg, Nucl. Phys. B435 (1995) 129.}
\lref\Witten{P.C. Argyres, M.R. Plesser, N. Seiberg and E. Witten, Nucl. Phys.
{B461} (1996) 71, hep-th/9511154.}
\lref\Vaughn{S.P. Martin and M.T. Vaughn, Phys. Lett. B318 (1993) 331.}
\lref\Banks{T. Banks and A. Zaks, Nucl. Phys. B196 (1982) 189.}
{\nopagenumbers
\rightline{DAMTP/98-23}
\rightline{hep-th/9804101}
\vskip 2truecm
\centerline {\bigbf Constructing a {\bigit c}-function for SUSY Gauge Theories}
\vskip 2.0 true cm
\centerline {D.Z. Freedman* and H. Osborn**\footnote{}{emails:
{\tt dzf@math.mit.edu} and {\tt ho@damtp.cam.ac.uk}}}
\vskip 12pt
\centerline {*\ Department of Mathematics and Center for Theoretical Physics,}
\centerline {MIT, Cambridge MA 02139, USA}
\vskip 8pt
\centerline {**\ Department of Applied Mathematics and Theoretical Physics,}
\centerline {Silver Street, Cambridge, CB3 9EW, England}
\vskip 2.0 true cm

{\eightpoint
\parindent 1.5cm{

{\narrower\smallskip\parindent 0pt
Recently a non-perturbative formula for the RG flow between UV and IR
fixed points of the coefficient in the trace of the energy momentum tensor
of the Euler density has been obtained for $N=1$ SUSY gauge
theories by relating the trace and {R-current} anomalies. This result
is compared here with an earlier perturbation theory analysis based on 
a naturally defined metric on the space of couplings for general 
renormalisable quantum field theories.  This approach is specialised to 
${N=1}$ supersymmetric theories and extended, using consistency
arguments, to obtain the Euler coefficient at fixed points to
4-loops. The result agrees completely, to this order, with the exact formula.

\narrower}}

\vfill
\eject}}
\pageno=1

The work of Seiberg \Sei\  and many other authors has greatly clarified the
non-perturbative structure of $N=1$ supersymmetric theories. There is 
considerable
evidence that there exist infrared attractive fixed points in many models,
where beta functions vanish and which define non-trivial superconformal
invariant theories.  As a consequence there has been renewed interest in the 
quest for a four-dimensional version of the Zamolodchikov $c$-theorem, 
which defines a function of the couplings for two-dimensional field theories
which decreases monotonically under RG flow to the infra
red limit and coincides at fixed points with the Virasoro central charge
of the associated conformal theory. 

In a four-dimensional theory coupled to a background metric, 
the external trace anomaly of the energy-momentum tensor is
\eqn\one{
 16\pi^2 \, T_{\alpha}{}^{\alpha} = \hbox{operator terms} +c \,
C^{\alpha\beta\gamma\de} C_{\alpha\beta\gamma\de}
- a \,  R^{\alpha\beta\gamma\delta} R^*{}_{\!\! \alpha\beta\gamma\delta} \, ,
}
where $C_{\alpha\beta\gamma\de}$ is the Weyl tensor and 
$R^*{}_{\!\! \alpha\beta}{}^{\gamma\delta} = {1\over 4}\ep_{\alpha\beta\ep\eta}
\ep^{\gamma\delta\si\rho}R^{\ep\eta}{}_{\si\rho}$, so that
$R^{\alpha\beta\gamma\delta} R^*{}_{\!\! \alpha\beta\gamma\delta}$ is the 
Euler density.
Cardy \Cardy\ has suggested that the Euler anomaly coefficient $a(g)$ as
a function of the running couplings $g^I(\mu)$
provides the desired $c$-function. The total flow $a_{UV}-a_{IR}$ between
UV and IR fixed points has been calculated
in both non-supersymmetric \Cap\ and more recently in 
many supersymmetric models \refs{\Bast,\Anone,\Antwo} and has been found to be 
positive in all models for flows from the trivial asymptotic freedom
UV fixed point, while
$c_{UV}-c_{IR}$ has no definite sign in the models studied.

In the analysis of \refs{\Anone,\Antwo} the values $a_{UV}$ and $c_{UV}$ were 
obtained from the free field content of the asymptotically free theories
studied, while $a_{IR}$ and $c_{IR}$ were related to the R-current anomalies
usually computed in studies of $N=1$ duality. At a fixed point, where the
renormalisation group beta functions vanish, by virtue of superconformal 
invariance the  R-charges of chiral fields are related to their scale dimensions
so that the result for the flow of $a$ can be simply written in terms of the 
anomalous dimensions $\gamma_r$, at the IR fixed point, of the chiral
superfields in the theory.
With $r$ denoting the gauge group representation $R_r$, of dimension
$\dim R_r$, the non-perturbative formula, for all models with a unique
R-current, is 
\eqn\fpa{
8(a_{UV}-a_{IR}) = \sum_r{\rm dim} R_r \, 
\gamma_r{}^{\!2}\big (1 - {\ts {2\over 3}} \gamma_r\big )\, ,
}
(Analogous formulae for the flows of $c$ and of flavour current
anomalies are given in \Antwo\ but they are not relevant here). 
The formula \fpa\ is a non-perturbative
result, but it is a desirable check of the methodology to compare it with
perturbative calculations which is the purpose of the present note. For
models with a perturbatively accessible fixed point, the $\gamma_r$ and
also the coefficient $a$ in \one\ can be
expressed as a power series in the couplings so that comparison can be made.

We now discuss the perturbative formalism of \refs{\JO\Weyl}, in which the
usual RG equations describing the variation
of an arbitrary renormalisation scale $\mu$ were extended to local
rescalings of the background space-time metric. In this approach $a$ can be
calculated to high orders of perturbation theory in a general renormalisable
quantum field theory, thus generalising earlier results in special cases
\Hath. (The present notation for
the anomalies in \one\ is related to that of \refs{\JO\Weyl} by  
$a =  16\pi^2 \beta_b  , \, c = - 16\pi^2 \beta_a$.) 
The initial study \JO\ used dimensional regularisation, but the results were
also shown \Weyl\ to follow from consistency conditions related to the 
commutativity of local rescalings. In a renormalisable theory with couplings
$g^I$ and associated beta functions $\beta^I$, it was shown how to define, to
all orders in perturbation theory, a metric $g_{IJ}=g_{JI}$ on the space of
couplings and associated one-form $W_I$ so that
\eqn\prta{
\pr_I \, 8{\tilde a} = t_{IJ}\beta^J \, , \qquad
t_{IJ} = g_{IJ} +  \pr_I W_J - \pr_J W_I \, , \qquad 
8{\tilde a} = 8 a + W_I \beta^I \, .
}
$g_{IJ}$ and also $W_I$ may be calculated in terms of connected one particle
irreducible vacuum graphs in the background geometry and \prta\ then permits
calculations of the perturbative expansion for $a$. The lowest order corrections
to the one loop free field results occur in general at three loops and were
given in \JO. If $W_I$ is an exact one-form then \prta\ represents a gradient
flow \grad. Clearly at a fixed point $\ta$ is equal to $a$
and under renormalisation flow \prta\ implies
\eqn\flow{
{\d \over \d t} \, g_t^I = - \beta^I(g_t^{\vphantom g}) \quad \Rightarrow
\quad {\d \over \d t}\, 8{\tilde a}(g_t^{\vphantom g}) = -
g_{IJ}(g_t^{\vphantom g})
\beta^I(g_t^{\vphantom g})\beta^J(g_t^{\vphantom g}) \, ,
}
Thus, if the metric is positive, ${\tilde a}(g_t)$
monotonically decreases, as does the Zamolodchikov $c$-function in two
dimensions. The positivity properties of the metric are not known in
general but the leading perturbative contribution is positive,
so one can establish the $c$-theorem at least in some region of weak coupling.
It should be noted that it is the associated quantity $\tilde a$ which is 
monotonic and not the anomaly $a$ itself (which actually increases as one 
flows away from the UV fixed point  in some models). Further the formula \fpa\ 
was derived under the condition that $\beta$-functions vanish, so our 
comparison can only be made at an IR fixed point.

In this note we apply the method of \refs{\JO,\Weyl}
to general $N=1$ supersymmetric
theories, assuming a simple gauge group $G$ with coupling $g$ and chiral
$G$-invariant couplings $Y_{ijk}, \ {\bar Y}^{ijk}$ so that the superpotential
is ${1\over 6}Y_{ijk}\phi^i\phi^j\phi^k$ for $\phi^i$ complex chiral superfields
acting on which the generators of $G$ are  $t_a = - t_a{}^{\!\dagger}$.
In an appendix the 3-loop results of \JO\ are briefly reviewed and
specialised to
the case where gauge, Yukawa, and quartic scalar couplings are related by
supersymmetry. This gives a straightforward check of the formula (2) to 
3-loop order. However the integrability conditions following from \prta\ 
allow, without any new Feynman graph calculations, the expression for the
metric to be extended to higher order and then 4-loop results for $\ta$ to be
found. To achieve this we use the  NSVZ beta function \NSVZ\ for the gauge
coupling $g$ to three loop order.

It is convenient to eliminate factors of $4\pi$ by writing
$\hg = g/4\pi, \ \hY_{ijk} = Y_{ijk}/4\pi , \ \hbY{}^{ijk} = 
\bY{}^{ijk}/4\pi$. We then restrict the results of two loop calculations
in \JO, as given in (A.1), to a supersymmetric theory with
$n_V$ vector superfields and $n_S$ chiral superfields. The metric and
one-form become
\eqn\metric{ \eqalign{
\d s^2 =
g_{IJ}(g)\d g^I \d g^J ={}& 4n_V \, {1\over \hg^2}(\d \hg)^2 (1 + \si \hg^2 )
+ {\textstyle{2\over 3}} \, \d {\hbY}{}^{ijk} \d \hY_{ijk} \, , \cr
W_I(g)\d g^I = {}& 2 n_V \, {1\over \hg}\d \hg (1 + \half \si \hg^2 ) +
{\textstyle{1\over 12}} \big ( {\hbY}{}^{ijk} \d \hY_{ijk} +
\d {\hbY}{}^{ijk} \hY_{ijk}\big )  \, . \cr}
}
The parameter $\si$, which is linear in $R$ and $C$ (where $\tr(t_a t_b)
= -R\de_{ab}$ and $C$ is similarly defined for the adjoint representation,
$C=N$ for $G=SU(N)$), is scheme dependent. To the order given in \metric\
the metric is clearly flat and the one form is exact. The leading
perturbative result is also manifestly positive which should therefore
remain valid at least in some region of weak coupling.
The corresponding 3-loop order result (A.5) for $a$, which is
independent of $\sigma$ because of cancellation between the
the $\ta$ and $W_I\beta^I$ terms of \prta, becomes
\eqn\athree{
8 a = 8 a_0 + 2n_V C(3C-R) \hg^4 - 4\, {{\rm tr}}(t^2 t^2) \hg^4
- {\rm tr}\big ( t^2 ({\hbY} \hY) \big ) \hg^2 \, , \qquad
({\hbY} \hY )^i{}_j = {\hbY}{}^{ilm}\hY_{jlm} \, ,
}
with $a_0$ the result for free fields, 
$8a_0 = {\textstyle{1\over 6}} (9n_V + n_S)$.
For asymptotically free theories
$a_0 = a_{UV}$, the value at the UV fixed point. Applying \athree\ to pure
SSYM theory without chiral matter gives $8a = {3\over 2}n_V ( 1+ 4C^2\hg^4)$
which increases away from the free UV fixed point. This shows that $a$
itself does not define a good $c$-function describing monotonic RG flow
between fixed points. The three loop corrections in both (A.5) and \athree\
are proportional to the two loop terms in the gauge beta function and are
scheme independent.

In this paper we require expressions for the gauge beta function to three
loops and for the Yukawa beta function to two loops. At this order they
are dependent on the choice of renormalisation scheme. For the gauge 
$\beta$-function we use the NSVZ form, which expresses $\beta^g$ to all
orders in terms of the anomalous dimension matrix $\gamma^i{}_j$ for the
chiral fields and
has been verified at three and four loops in \JJN. This is here written as
\eqn\betag{
{1\over g}\beta^g (g) = {\tbe(\hg) \over 1-2C \hg^2} \, , \qquad
\tbe(\hg) = - \big (\beta_0 - 2\, {\overline{\rm tr}}(\gamma t^2)\big)\hg^2 \, ,
\quad \beta_0 = 3C-R \, ,
}
where ${\overline{\rm tr}} = n_V^{-1}{\rm tr}$. The Yukawa beta function
is also in general directly expressible in terms of $\gamma$, as a 
consequence of the non-renormalisation of the superpotential,
\eqn\Yuk{
\beta^Y{}_{\!\! ijk} = 3 \, Y_{\ell(ij}\gamma^\ell{}_{k)} \, , \qquad
{\bar \beta}^{\bY ijk} = 3 \, \gamma^{(i}{}_{\ell}\bY^{jk)\ell} \, .
}
To one and two loop order using dimensional reduction \JJ\foot{The 
compatibility of these results
with the NSVZ $\beta$-function with independent 3 loop calculations for 
$\beta^g$, up to the freedom corresponding to a change of scheme,
was shown in \refs{\JJN,\JJ}. The form assumed \Yuk\ does
not itself remove possible scheme dependence since
taking $\de Y_{ijk} = 3Y_{\ell(ij}h^\ell{}_{k)}, \
\de \bY^{ijk} = 3h^{(i}{}_{\ell}\bY^{jk)\ell}$ and using the general result 
for the variation of the $\beta$-function, 
$\de \beta^I = \beta^J\pr_J \de g^I - \de g^J \pr_J \beta^I$, gives in \Yuk\
$\de \gamma = \beta{\cdot \pr}\, h - {\de g}{\cdot \pr}\,\gamma 
+ [\gamma, h]$. At a fixed point then $\de \gamma_* = [\gamma_* , h_*]$ 
so that the eigenvalues of $\gamma_*$ are invariant.
However this does not maintain the NSVZ form in \betag\ which appears to
define the renormalisation scheme uniquely.}
\eqn\gg{\eqalign{
\gamma^{(1)i}{}_j = {}& P^i{}_j \equiv {\textstyle{1\over 2}}
({\hbY} \hY )^i{}_j + 2\hg^2 (t^2)^i{}_j \, , \cr 
\gamma^{(2)i}{}_j = {}& - \hbY{}^{ik\ell}\hY_{jkm} P^m{}_{\!\ell} 
+ 2\hg^2 (Pt^2)^i{}_j + 2\beta_0 \hg^4 (t^2)^i{}_j \, . \cr}
}
{}From (A.9) in the same dimensional reduction scheme
\eqn\con{
\qquad \si = -2C + 5\beta_0 \, .
}

We first show that \athree\ at a perturbative fixed point 
agrees with \fpa\ to 3-loop order. Using ${\hbY}{\cdot \beta^\hY} =
3\, \tr((\hbY \hY) \gamma)$ and the one loop form for $\gamma$ in \gg\
it is easy to see that
\eqn\afp{
8a = 8a_0  - {\rm tr} ( \gamma \gamma)
+2 n_V C\beta_0 \hg^4 + {\textstyle{1\over 6}}\, {\hbY}{\cdot \beta^\hY} \, .
}
To obtain a perturbative fixed point it is necessary to choose the gauge
group $G$ and representation content of other fields so that $\beta_0$ is
effectively small. This is possible  \Banks\ in a large $N$ limit, where as 
$N\to \infty$, the gauge group  is such that $C= {\rm O}(N)$, 
$n_V= {\rm O}(N^2)$, and  $\beta_0>0$ is
tuned to be ${\rm O}(1)$. In this case there is a IR fixed point with
$g_*^{\, 2} = {\rm O}(N^{-2})$ and the other couplings
are given in terms of $g_*$ so that in the case of $N=1$ SUSY theories
discussed above $\bY_*, Y_* = {\rm O}(g_*)$.
In these circumstances $\gamma =  {\rm O}(N^{-1})$ and
${\rm tr} ( \gamma \gamma) = {\rm O}(1)$ while the last two terms in
\afp\ are of higher order in $1/N$. It is evident therefore 
that  \afp\ is compatible with \fpa\ to leading order.
Equivalently in discussing a
perturbative fixed point it is legitimate to assume that
the lowest order beta functions which actually appear in \afp\ are `completed'
by higher order terms, $-\beta_0 \hg^2 \to \tbe(\hg)(1+{\rm O}(\hg^2))$,
which make them vanish. This allows the last
two terms in \afp\ to be dropped and we then observe agreement with 
\fpa\ to 3-loop order.
{\midinsert
\hfil \epsfxsize=0.9\hsize
\epsfbox{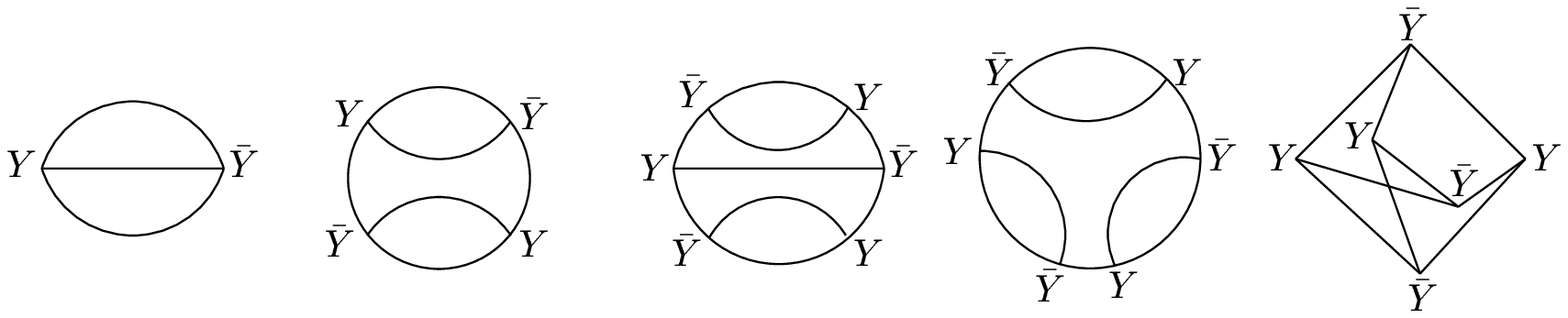} \hfil
\vskip -6pt
{\eightpoint{
\parindent 1.5cm

{\narrower\smallskip\noindent
Fig. 1 Two, three and four loop vacuum diagrams for chiral scalar
fields. Each diagram represents a potential contribution to $a$, although
the results here contain no term corresponding to the non planar diagram, while
selecting one vertex determines the possible forms for the one form $W$
and pairs of vertices the metric $g$. Since there is just a single diagram at
two and three loops $W$ is exact to this order while there are three
possible terms in the metric at three loops.
\smallskip}

\narrower}}
\endinsert}
We now demonstrate that the formalism can be extended in a fairly simple way to
obtain results for $\ta$ to four-loop order, which can then be compared with
the exact formula \fpa. The main idea is to use the integrability conditions
for \prta\ to generate the three-loop contributions to $g_{IJ}$ and $W_I$ in 
the scheme in which the beta functions are presented above. 
We begin with the the
$g_{Y\bY}$ sector and then extend results to the complete metric.
The form of possible 
contributions are determined by vacuum diagrams, exhibited in Fig. 1 for
the cubic interactions determined by the superpotential. 
Hence, for suitable coefficients $\alpha,\beta,\gamma, \epsilon$ (which in 
general are scheme dependent\foot{If, in terms of the previous footnote,
we take $h=\lambda (\hbY \hY)$ then $\de \alpha = \de \beta = -4\lambda, \
\de \gamma = - 2 \lambda$, while $\de \gamma^{(2)i}{}_j = 4\lambda \,
\hg^2 ((\hbY \hY) t^2 )^i{}_j + 8\lambda\, \hg^2 
\hbY{}^{ik\ell}\hY_{jkm} (t^2)^m{}_{\!\ell}$}), it may be written as
\eqn\metricY{\eqalign{
\d Y{\cdot g_{Y\bY}}{\cdot \d\bY} = {}& {\textstyle{2\over 3}} \, 
\d {\hbY}{}^{ijk} \d \hY_{ijk}
+ \alpha \, \tr \big ( (\d\hbY \d\hY)(\hbY \hY) \big )
+ \beta\, \tr \big ( (\d\hbY \hY)(\hbY \d\hY) \big ) \cr
{}& + \gamma\,  \tr \big ( (\hbY \d\hY)(\hbY \d\hY) + 
(\d\hbY \hY)(\d\hbY \hY) \big )
+ \epsilon\, \hg^2 \tr \big ( (\d\hbY \d\hY) t^2 \big ) \, . \cr}
}
In a similar fashion we may extend the result in \metric\ for 
${\bar W}_Y{\cdot \d Y} + W_\bY{\cdot \d \bY}$ to allow for 3 loop 
contributions. However it remains exact to this order and may be
disregarded so that $t_{Y\bY}=t_{\bY Y} = g_{Y\bY}$. At the same effective
order we may also take
\eqn\tyg{
\d Y{\cdot t_{Yg}}\,\d g + \d \bY{\cdot t_{\bY g}}\,\d g = 2\eta \, \hg\d \hg\,
\tr \big ( \d(\hbY \hY) t^2 \big ) \, .
}
Using \metricY\  and \tyg\ with the the results for
$\beta^Y$ provided by \Yuk\ and \gg\ then gives an equation determining
the dependence of $\ta$ on $\bY$,
\eqnn\prY
$$\eqalignno{\!\!\!
\d \bY{\cdot \pr_\bY} 8\ta& = \half \tr \big ( (\d\hbY \hY)
(\hbY \hY) \big ) + 2\hg^2 \tr \big ( (\d\hbY \hY) t^2 \big ) \cr
{}&+ \half \alpha \, 
\d \hbY{}^{ik\ell}\hY_{jkm} (\hbY \hY)^m{}_{\!\ell} (\hbY \hY)^j{}_i
+ \half(\beta+2\gamma-1) \hbY{}^{ik\ell}\hY_{jkm} (\d \hbY \hY)^m{}_{\!\ell}
(\hbY \hY)^j{}_i \cr
{}& + \quar (\alpha + \beta + 2\gamma)
\tr \big ( (\d \hbY \hY) (\hbY \hY) (\hbY \hY) \big ) 
+ (2\alpha + \half \epsilon)\hg^2 
\d \hbY{}^{ik\ell}\hY_{jkm} (\hbY \hY)^m{}_{\!\ell}
(t^2)^j{}_i \cr
{}& + 4 \alpha \, \hg^2 \hbY{}^{ik\ell}\hY_{jkm} (\d \hbY \hY)^m{}_{\!\ell}
(t^2)^j{}_i + (3\alpha + \quar \epsilon + 2)\hg^2
\tr \big ( (\d \hbY \hY) (\hbY \hY) t^2 \big ) \cr
{}& + 2\epsilon\, \hg^4 \d \hbY{}^{ik\ell}\hY_{jkm}(t^2)^m{}_{\!\ell}(t^2)^j{}_i
+(\epsilon+4) \hg^4 \tr \big ( (\d \hbY \hY)  t^2 t^2 \big )\cr
{}& + 2 (1-\eta) \beta_0 \, \hg^4 \tr \big ( (\d \hbY \hY)  t^2 \big ) \, . 
& \prY \cr}
$$
The first line here follows just from \metric\ and the one loop result in
\gg. Integrability of \prY\ imposes the conditions 
\eqn\inta{
\beta+2\gamma-1 = 2\alpha \, , \qquad \epsilon = 4 \alpha \, ,
}
and then  \prY\ then gives
\eqn\aY{\eqalign{
8\ta = {}& 8a_0 + \quar \tr \big ( (\hbY \hY) (\hbY \hY) \big ) 
+ 2\hg^2 \tr \big ((\hbY\hY)t^2\big )\big ( 1+(1-\eta)\beta_0 \hg^2 \big ) \cr 
{}& + 2\alpha \, \hbY{}^{ik\ell}\hY_{jkm} P^m{}_{\!\ell} P^j{}_i 
+ {\ts{1\over 12}}(3\alpha+1)\tr \big ( (\hbY \hY)(\hbY\hY) (\hbY \hY)\big ) \cr
{}& + (2\alpha+1)\hg^2 \tr \big ( (\hbY \hY) (\hbY \hY) t^2  \big )
+ 4(\alpha+1)\hg^4 \tr \big ( (\hbY \hY) t^2 t^2 \big ) + \dots \,.
\cr}
}

It remains to consider also the dependence on the gauge coupling. To this
end we extend the metric in \metric\ to
the form
\eqn\metricg{
g_{gg}\d g^2 = 4n_V \, {1\over \hg^2}(\d \hg)^2 \Big (1 + \si \hg^2  + \tau
\hg^2 {\overline{\rm tr}} \big ( (\hbY \hY) t^2 \big ) + \rO(\hg^4) \Big )\, ,
}
as well as requiring
\eqn\tgy{
\d g\, t_{gY}{\cdot \d Y} + \d g \,t_{g\bY}{\cdot \d \bY} = 
2\kappa \, \hg\d \hg \, \tr \big ( \d(\hbY \hY) t^2 \big ) \, .
}
Then \prta, with \betag, gives
\eqn\prg{\eqalign{
g \pr_g 8\ta ={}& 4n_V {\tilde \beta}(\hg)  \big (1 + (2C+\si)\hg^2 + \tau
\hg^2 {\overline{\rm tr}} \big ( (\hbY \hY) t^2 \big ) + \rO(\hg^4) \big )\cr
{}& + 2\kappa \, \hg^2 \tr \big ( ({\bar \beta}^\hbY \hY ) t^2 + 
(\hbY \beta^\hY) t^2 \big ) \, . \cr}
}
It is of course crucial that the $Y,\bY$ dependent terms obtained from
integrating \prg\ are consistent with those already found \aY. This requires,
after using \con,
\eqn\consist{
\kappa - 2\alpha = 1 \, , \qquad \tau - 2\eta = 3 \, .
}
We may then calculate the purely $\hg$ dependent
terms in $\ta$ from \prg\ to be
\eqn\tag{
8\ta = { -2}n_V \beta_0 \hg^2\big (1 + {\ts {5\over 2}} \beta_0\hg^2\big )
+ 4\hg^4 \tr ( t^2 t^2) + {\ts {16\over 3}}\, \hg^6
\tr ( t^2 t^2 t^2 ) +  \rO( \beta_0\hg^6) \, .
}
Combining \aY\ and \tag\ we can express $\ta$ to effectively 4 loop order
in the form
\eqn\taf{ \eqalign{
8\ta = {}& 8a_0  -2n_V \beta_0 \hg^2\big (1+{\ts {5\over 2}} \beta_0\hg^2\big )
+ \tr ( PP ) \cr
{}& +2\alpha\, \hbY{}^{ik\ell}\hY_{jkm} P^m{}_{\!\ell} P^j{}_i 
+ {\ts{2\over 3}}(3\alpha+1) \, \tr(PPP) -4\alpha \,\hg^2 \tr(PPt^2)\cr
{}& +  \rO( \beta_0\hg^6) + \rO( \beta_0\hg^4) \tr(Pt^2)  \, . \cr}
}
For finite theories with zero beta functions $\ta = a_0$ is
independent of the couplings and this is reflected in \taf\ since
the corrections vanish if  $\beta_0=0, \, P=0$ which are sufficient
for a finite theory to two loop order.

To make the connection with the results in \Antwo\ we use, to the required
order,
\eqn\form{\eqalign{ \!\!\!\!\!
n_V {\tilde \beta}(\hg) + {\ts{1\over 6}}\, \hbY{\cdot \beta}^{\hY} \!
={}&{ - n_V} \beta_0 \hg^2 + \tr(PP) 
- \hbY{}^{ik\ell}\hY_{jkm} P^m{}_{\!\ell} P^j{}_i \cr
{}& + 2\hg^2 \tr(PPt^2) + 2\beta_0 \hg^2 \tr(Pt^2) \, , \cr
\tr (\gamma\gamma) ={}& \tr(PP) - 
2\, \hbY{}^{ik\ell}\hY_{jkm} P^m{}_{\!\ell} P^j{}_i
+ 4\hg^2 \tr(PPt^2) \, , \cr}
}
and hence \taf\ becomes consistent with
\eqn\aa{\eqalign{
8\ta = {}& 8a_0  - \tr(\gamma\gamma) + {\ts{2\over 3}}\,
\tr(\gamma\gamma\gamma) \cr
{}& + 2n_V {\tilde \beta}(\hg)\big (1+{\rm O}(\beta_0 \hg^2)\big )
 + {\ts{1\over 3}}\, \hbY{\cdot \beta}^{\hY}
+\alpha  \, \tr \big((\hbY \beta^{\hY})P\big ) \, .\cr}
}
At a fixed point the second line, and hence the dependence on the
undetermined parameter $\alpha$, vanishes. The result \aa\ is in accord with
\fpa\  taking $\gamma$ at the IR fixed point to have eigenvalues $\gamma_r$. 
It is a tribute to the consistency of quantum field
theory that two such different approaches give identical results---
the non-perturbative method of \Antwo\ effectively uses only 1-loop
triangle anomalies while the second method strictly requires 4-loop
calculations in curved space which were bypassed here by use of the
consistency conditions associated with \prta.

As a specific illustration we consider an example based on  magnetic SQCD
\Sei, as described by Kogan {\it et al} \Kogan, 
which for $G=SU(N)$ contains chiral
fields $\varphi^i, \, {\tilde \varphi}_i$ belonging to $N, \, {\overline N}$
representations of $SU(N)$, with $i=1, \dots N_f$ as well as a $SU(N)$
singlet $M^i{}_j$. There are then Yukawa interactions preserving
$SU(N_f)\times SU(N_f)$ symmetry represented by the superpotential 
$y\, {\tilde \varphi}_i M^i{}_j \varphi^j$ and the matrix $P$ defined in \gg\
has the form, in a basis for the fields given by $(M,\varphi,{\tilde \varphi})$,
\eqn\PP{
P = \pmatrix{ N \hy^2 \b1_{N_f^2} & 0 & 0\cr
0& \big ( N_f \hy^2 - {\ts{N^2-1 \over N}}\hg^2 \big ) \b1_{NN_f} & 0\cr
 0 & 0 & \big ( N_f \hy^2 - {N^2-1 \over N}\hg^2 \big ) \b1_{NN_f} } \, .
}
For this theory $\beta_0=3N-N_f$ and the lowest order formula for $\ta$
becomes
\eqn\Ka{ \eqalign{
8 \ta ={}& {\ts{1\over 6}} \big (9(N^2-1) + 2NN_f \big )\cr
{}&  - 2(N^2-1)(3N-N_f)\hg^2 + N^2 N_f{}^{\!2} \hy^4 + 
2{N_f\over N} \big ( NN_f \hy^2 - (N^2-1) \hg^2 \big )^2 \,. \cr}
}
This satisfies, to lowest order since from \metric\ $\d s^2 = 4(N^2-1)
\d \hg^2/\hg^2 + 4 N N_f \d \hy^2$,
\eqn\betayg{
\hg{\pr\over \pr \hg}8\ta = 4(N^2-1) {\tilde \beta}^g(\hg, \hy) \, , \qquad
{\pr\over \pr \hy}8\ta = 4NN_f{}^{\!2} \beta^\hy(\hg,\hy) \, ,
}
and has a minimum at the IR fixed point
\eqn\fp{
NN_f \hy_*^2 = {2N_f \over N+2N_f} \, (N^2-1) \hg_*^2 = 3N-N_f \, .
}
It is easy to show that then
\eqn\fpp{
8(a_{UV}-a_{IR}) = {\rm tr}(P_*^2)= {N+2N_f \over 2N_f}\, (3N-N_f)^2 \, .
}
Of course these perturbative results are only valid in a large $N$ limit
with $N_f$ tuned to ensure $3N-N_f = {\rm O}(1)$.

We may also note the restriction to $N=2$ SUSY theories which may be
defined by taking the superpotential to be $\sqrt 2 g\, \eta_a \chi^T  T_a \xi$
where $\eta, \xi$ and $\chi$ are chiral superfields transforming according
to the adjoint, $R$ and $R^*$ representations of the gauge group. In this
case there is a single coupling $g$ with the perturbative beta function
non zero only at one loop with $\beta_0 = 2(C-R)$ where $\tr(T_aT_b)= -R
\delta_{ab}$. From \metric\ and \con\ we have
\eqn\metrictwo{
g_{gg}\d g^2 = 4n_V \, {1\over \hg^2}(\d \hg)^2 \Big (1 + 4\beta_0 \hg^2  
\Big )\, ,
}
and, with the analogous result for $W$, $a=a_0$ to three loops, independent
of $g$. The finiteness of $N=2$ theories beyond one loop, in particular
the vanishing of $\gamma$, suggest that this
may be true to all orders, compatible with there being  no perturbative fixed
point.

\bigskip
\appendix{A}{}
In \JO\ results were described for the non supersymmetric theories.
In order to show the connection with above results for SUSY models
we describe briefly the necessary transcription code. For $n_\phi$
real scalars $\phi^i$ and $n_\psi$ Majorana fermions $\psi$, ${\bar \psi} =
\psi^T C, \ C= -C^T$, the essential 
dimensionless couplings for a general renormalisable four dimensional theory
are $g^I=(g,\Gamma_i,\lambda_{ijk\ell})$, with $g$ the gauge coupling,
a Yukawa interaction ${1\over 2} {\bar \psi}\Gamma_i \psi \phi^i$, where
$\Gamma_i$ is a matrix linear in $\gamma_5$ satisfying $C \Gamma_i = -
( C \Gamma_i )^T$ (${\hat \Gamma}_i$ is also defined by $\gamma_\mu \Gamma_i = 
{\hat \Gamma}_i \gamma_\mu$), and a quartic scalar interaction 
${1\over 4!}\lambda_{ijk\ell}\phi^i\phi^j\phi^k\phi^\ell$. The gauge
interactions are specified by the gauge group generators
$t_a^\phi , \, t_a^\psi$, $ a=1,\dots n_V$ acting on $\phi,\, \psi$ satisfying
$t_a^\phi = - t_a^{\phi T}$, $C t_a^\psi C^{-1} = - {\hat t}_a^{\psi T}$
(where $\gamma_\mu t_a^\psi = {\hat t}_a^{\psi} \gamma_\mu$). For such couplings
the metric calculated in \JO\ can be written as
\eqn\metg{
\d s^2 = 4n_V \, {1\over g^2}(\d g)^2 \bigg (1+ A\, {g^2\over16\pi^2} \bigg ) 
+ {1\over 16\pi^2} \, {\textstyle{1\over 6}}\, {\rm tr} \big (
\d {\hat \Gamma}_i \d \Gamma_i \big )
+ {1\over (16\pi^2)^2} \, {\textstyle{1\over 72}} \, \d \lambda_{ijk\ell}
\d \lambda_{ijk\ell} \, .
}
The associated one form to the same order was also
\eqn\Wg{
W_I(g) \d g^I = 2n_V \, {1\over g}\d g \bigg(
1 + {\textstyle{1\over 2}} A\, {g^2\over 16\pi^2}  \bigg )
+ {1\over 16\pi^2} \, {\textstyle{1\over 24}}\, {\rm tr} \big (
{\hat \Gamma}_i \d \Gamma_i \big ) + {1\over (16\pi^2)^2} \,
{\textstyle{1\over 216}} \, \lambda_{ijk\ell} \d \lambda_{ijk\ell} \, .
}
The trace in the terms involving $\Gamma$, and $t^\psi$,
 also involves a sum over spinor
indices. Using the results for the $\beta$-functions
\eqn\betag{\eqalign{
{1\over g}\beta^g  = {}&  - \beta_0 {g^2 \over  16\pi^2}  -  
{g^2  \over  (16\pi^2)^2} \Big ( \beta_1 g^2 - 
{\textstyle{1\over 4}}\, {\overline{\rm tr}}
\big ( t^{\psi \, 2} {\hat \Gamma}_i \Gamma_i ) \Big ) \, , \cr
\beta^\Gamma{}_{\!\! i} = {}& {1\over  16\pi^2} \Big (
\Gamma_j {\hat \Gamma}_i \Gamma_j + \half \big (
\Gamma_j {\hat \Gamma}_j + 6 {\hat t}^{\psi \, 2} \big ) \Gamma_i
+ \Gamma_i \half \big ( {\hat \Gamma}_j \Gamma_j + 6 t^{\psi \, 2} \big )
+{\ts {1\over 4}}\, \tr ( \Gamma_i  {\hat \Gamma}_j )  \Gamma_j \Big )
\, , \cr}
}
where
\eqn\betae{\eqalign{
\beta_0 = {}& {\textstyle{1\over 3}} \big ( 11 C - 2 R_\psi -
{\textstyle{1\over 2}} R_\phi \big ) \, , \quad R_\phi = - {\overline{\rm tr}}
\big ( t^{\phi \, 2} \big )\, , \  R_\psi = - {\textstyle{1\over 4}} \,
{\overline{\rm tr}} \big ( t^{\psi \, 2} \big ) \, , \cr
\beta_1 = {}& {\textstyle{1\over 3}} C \big ( 34 C - 10 R_\psi - R_\phi \big )
- {\textstyle{1\over 2}}\, {\overline{\rm tr}} \big ( t^{\psi \, 2}
 t^{\psi \, 2} \big ) - 2\, {\overline{\rm tr}}
\big ( t^{\phi \, 2} t^{\phi \, 2} \big )\, , \cr}
}
then $a$ can be found, from \prta\ and the results for free fields,
to three loops, independent of $A$, to be
\eqn\ag{
8a =  {\textstyle{1\over 90}} \big (124n_V + 11 n_{\psi} + 2 n_\phi \big )
+{1\over (16\pi^2)^2}\Big (  n_V \beta_1 g^4
-  {\textstyle{1\over 4}}\, {\rm tr}\big ( t^{\psi \, 2} {\hat \Gamma}_i
\Gamma_i ) \, g^2 \Big ) \, .
}
Completing the calculation in \JO\ to include scalar fields, using
dimensional regularisation, gives
\eqn\dimreg{
A_{\rm dim. reg.} = 17 C - {\textstyle{10\over 3}} R_\psi 
 - {\textstyle{7\over 6}} R_\phi \, .
}

The reduction to a general SUSY theory, as described above when
$n_\psi=n_V+n_S, \ n_\phi = 2n_S$, may be obtained by taking
\eqn\gen{
t_a^\phi \to \pmatrix{t_a & 0\cr 0 & - t_a{}^{\!T}} \, , \qquad
{t_a^\psi \atop {\hat t}_a^\psi} \to \pmatrix{t_a{}^{\!{\rm ad}}  & 0\cr 0 &
t_a } P_\pm + \pmatrix{t_a{}^{\!{\rm ad}}  & 0\cr 0 & - t_a{}^{\!T}} P_\mp \, ,
}
with $t_a{}^{\!{\rm ad}}$ the adjoint representation generators and
$P_\pm = {1\over 2}( 1\pm \gamma_5)$, and also
\eqn\yuk{
{\Gamma_i \atop {\hat \Gamma}_i} \to \pmatrix{ 0 & -\sqrt 2 g (t_a)^i{}_{\! k}
\cr -\sqrt 2 g ( t_b{}^{\! T})_j{}^i & Y_{ijk}} P_\pm +
\pmatrix{ 0 & \sqrt 2 g (t_a{}^{\! T})_i{}^k \cr \sqrt 2 g (t_b)^j{}_i &
{\bar Y}^{ijk}} P_\mp \, .
}
Thus $R_\psi = C+R, \ R_\phi = 2R $ and \metg,\Wg\ reduce to \metric\ as well
as \ag\ leading to \athree. To apply \dimreg\ we transform to
dimensional reduction, as appropriate for supersymmetric perturbative
calculations, using \Vaughn\ $16\pi^2/g^2_{\rm reg.} = 16\pi^2/g^2_{\rm red.}
+ {1\over 3} C$ which gives
\eqn\dimred{
A_{\rm dim. red.} = A_{\rm dim. reg.} - {\ts{2\over 3}} C =
{\ts{1\over 3}}(39 C - 17 R ) \, , \quad \si = A_{\rm dim. red.} +
{\ts{2\over 3}} R = 13 C - 5R \, .
}

\bigskip
\noindent{\bigbf Acknowledgements}
\medskip
The research of D.Z.F. is supported in part by NSF Grant No. PHY-97-22072.
Both of us are very grateful to Ian Jack for several very helpful discussions.
\listrefs
\bye